\documentclass[10pt, conference,a4paper]{IEEEtran}
\makeatletter
\long\def\@makecaption#1#2{\ifx\@captype\@IEEEtablestring%
\footnotesize\begin{center}{\normalfont\footnotesize #1}\\
{\normalfont\footnotesize\scshape #2}\end{center}%
\@IEEEtablecaptionsepspace
\else
\@IEEEfigurecaptionsepspace
\setbox\@tempboxa\hbox{\normalfont\footnotesize {#1.}~~ #2}%
\ifdim \wd\@tempboxa >\hsize%
\setbox\@tempboxa\hbox{\normalfont\footnotesize {#1.}~~ }%
\parbox[t]{\hsize}{\normalfont\footnotesize \noindent\unhbox\@tempboxa#2}%
\else
\hbox to\hsize{\normalfont\footnotesize\hfil\box\@tempboxa\hfil}\fi\fi}

\usepackage[compatibility=false]{caption}
\DeclareCaptionFont{quackfont}{\fontfamily{ptm}\fontsize{7.5pt}{9pt}\selectfont}
\usepackage[labelformat=simple,font=quackfont]{subcaption}

\makeatother
\usepackage{lipsum}
\makeatletter
\let\MYcaption\@makecaption
\usepackage{cite}
\usepackage{graphicx}
\usepackage{subcaption}
\usepackage{float}
\usepackage{refstyle}
\usepackage{multicol, blindtext}
\usepackage{amsmath,amssymb,amsfonts}
\usepackage{adjustbox}
\newcommand{\norm}[1]{\left\lVert#1\right\rVert}

\usepackage[font=scriptsize]{caption}
\usepackage[font=footnotesize]{caption}
\usepackage{stfloats}
\usepackage{lmodern,bm}                
\usepackage[T1]{sansmath}
\usepackage{xcolor}
\usepackage{algorithm, algorithmic}
\usepackage{enumitem}
\usepackage{dutchcal}

\begin{document}

\title{An Unsupervised Learning-Based Approach for Symbol-Level-Precoding}

\author{\IEEEauthorblockN{Abdullahi Mohammad\IEEEauthorrefmark{1}, Christos Masouros\IEEEauthorrefmark{1} and Yiannis Andreopoulos\IEEEauthorrefmark{1}}\\
\IEEEauthorblockA{\IEEEauthorrefmark{1}Department of Electronic and Electrical Engineering, University College London, WC1E 7JE UK}\\[-2.0ex]% <-this % stops an unwanted space
e-mail: (abdullahi.mohammad.16; c.masouros; i.andreopoulos)@ucl.ac.uk
}

\maketitle

\begin{abstract}
This paper proposes an unsupervised learning-based precoding framework that trains deep neural networks (DNNs) with no target labels by unfolding an interior point method (IPM) proximal \textit{`log'} barrier function. The proximal \textit{`log'} barrier function is derived from the strict power minimization formulation subject to signal-to-interference-plus-noise ratio (SINR) constraint. The proposed scheme exploits the known interference via symbol-level precoding (SLP) to minimize the transmit power and is named strict Symbol-Level-Precoding deep network (SLP-SDNet). The results show that SLP-SDNet outperforms the conventional block-level-precoding (Conventional BLP) scheme while achieving near-optimal performance faster than the SLP optimization-based approach. 
\end{abstract}

\IEEEpeerreviewmaketitle

\section{Introduction}
\IEEEPARstart{R}{ecent} studies on interference exploitation have shown that known inferences can be effectively managed and transformed into valuable signals to improve the system's quality-of-service (QoS) \cite{masouros2007novel}. The concept of constructive interference (CI) is first introduced in \cite{masouros2007novel}, where instantaneous interference is categorized into constructive and destructive. Traditionally, multi-user interference (MUI) is suppressed in block-level precoding designs. However, the symbol-level precoding (SLP) technique utilizes the transmitted symbol to convert the  MUI into useful signals \cite{4801492}. Suboptimal strategies that exploit CI are first introduced in \cite{4801492}. Optimal SLP schemes using convex optimization-based CI with strict phase constraints on the received constellation point are proposed in \cite{masouros2015exploiting,alodeh2015constructive,masouros2018harvesting}. Despite the performance benefits offered by the optimization-based SLP schemes, computational complexity is still an issue in their implementation on practical systems.\par  
Due to the low computational cost of online training, there has been an increasing interest in designing deep learning (DL)  precoding schemes recently for MU-MISO downlink transmission \cite{alkhateeb2018deep,de2018robust,huang2018unsupervised,xia2019deep,sohrabi2020robust,9468383}. For example, in \cite{alkhateeb2018deep} the authors propose a DL-based coordinated beamforming technique to improve the link reliability and low latency in millimeter-wave (mmWave) communications. A deep neural network (DNN) precoding method is introduced in \cite{de2018robust} for decentralized decision making. An unsupervised learning (UL) based beamforming scheme that explores the optimal solution of weighted-sum-rate is proposed in \cite{huang2018unsupervised}. Convolutional neural networks (CNNs) framework for downlink beamforming optimization using expert knowledge based on the known structure of optimal iterative solutions is investigated in \cite{xia2019deep}. The application of DL on SLP is further studied in \cite{sohrabi2020robust}, where a deep autoencoder-based framework is designed for robust SLP and symbol detection. While the approach has low computational complexity, decision rules for symbol detection at the receivers are practically challenging to implement. To further enhance training efficiency, a computationally low-cost DNN-based SLP design is proposed in \cite{9468383}.\par
However, most of the learning-based strategies mentioned above are based on supervised learning, where the constraints are implicitly contained in the training dataset obtained from conventional optimization solutions. This requires solving the optimization problem twice, first by traditional optimization and second by approximating the optimal solution using DNN. However, if it was difficult to obtain the optimal solutions via conventional optimization methods, the learning-based solutions may be impractical.\par 
This paper proposes a learning-based precoding scheme that requires no target labels for power minimization problems under signal-to-interference-noise-ratio (SINR). The learning framework is designed by unfolding an IPM iterative algorithm via IPM proximal log barrier function that considers the convexity of the inequality constraint. A case scenario of strict phase angle rotation is considered under a known perfect channel condition.
\section{System Model}\label{section2}
Suppose a MISO downlink channel in a single cell with $N$ transmit antennas at the base station (BS) serves $K$ single-antenna users. The channel between users and the BS is assumed to be quasi-static flat-fading and is denoted by $\mathbf{h}_{i}\in \mathbb{C}^{{N}\times1}$.
\subsection{Conventional Power Minimization}
Traditionally, the power minimization problem tries to minimize the average transmit power by handling all interference as harmful subject to QoS constraints, as described below \cite{bjornson2014optimal}
\begin{equation} \label{conv_power_min}
    \begin{aligned}
    & \underset{\mathbf{\{v_{i}\}}}{\text{min}}
    & & {\sum_{i=1}^K\norm{\mathbf{v}_{k}}_{2}^2} \\
    & \text{s.t.}
    & &  \frac{|\mathbf{h}_{i}^{T}\mathbf{v}_{i}|^{2}}{\sum_{k=1,k\neq i}|\mathbf{h}_{i}^{T}\mathbf{v}_{k}|^{2}+{N}_{0}}\geq\Gamma_{i}\ , \ \forall {i}.
    \end{aligned}
\end{equation}
where $\Gamma_{i}$ is the SINR of the \textit{i-th} user. From an instantaneous viewpoint, problem (\ref{conv_power_min}) does not consider the fact that interference can additively improve the received signal power \cite{masouros2010correlation}. Therefore, the solution is sub-optimal.

\begin{figure}[!t]
    \includegraphics[width=2.8in,height=2.0in]{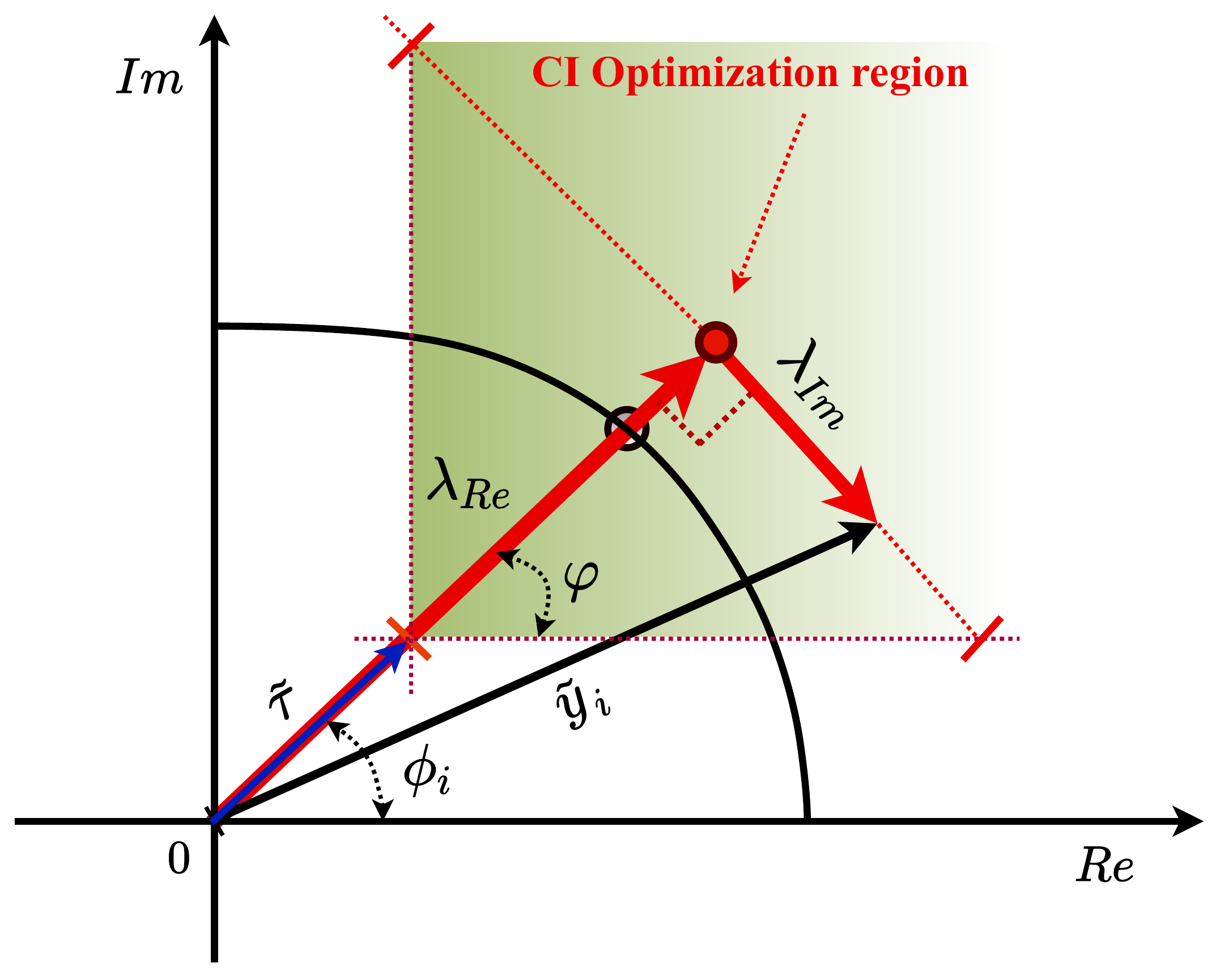}
    \caption{Generic geometrical optimization regions for interference exploitation for Precoding design in QPSK \cite{masouros2015exploiting}}
    \label{fig:CI_GEOMETRY}
\end{figure}

Fig. \ref{fig:CI_GEOMETRY} shows the the generic geometrical representation of the CI. The real part of the received symbol $(\lambda_{Re})$ gives a measure of the received constellation along the theoretical constellation axis. Likewise, the imaginary part $(\lambda_{Im})$ shows the extent of the phase displacement from the primary constellation point.

\subsection{Power Minimization via Symbol-Level Precoding}
The instantaneous interference in a multi-user downlink channel scenario for M array phase shift keying (M-PSK) modulation can be categorized into constructive and destructive based on the known criteria defined in \cite{masouros2013known}.
Therefore, CI is defined as the interference that pushes the received symbols away from the modulated-symbol constellation's decision edges \cite{masouros2015exploiting}. For further details on SLP and its formulation (3), we refer the reader to \cite{li2020tutorial}. Therefore, the problem in (\ref{conv_power_min}) is modeled to incorporate CI in the power minimization formulation. Consequently, the interfering signals align with the symbol of interest constructively by precoding vectors, offering useful signals. Hence, for M-PSK, the power minimization SLP-based optimization can be reformulated based on the classification criteria explained in \cite{masouros2015exploiting} 
\begin{equation} \label{p1}
    \begin{aligned}
    & \underset{\mathbf{\{v_{i}\}}}{\text{min}}
    & & {\norm{\sum_{k=1}^K \mathbf{v}_{k}{e}^{j(\phi_{k}-\phi_{1})} }_{2}^2} \\
    & \text{s.t.}
    & & \text{Im}{\left(\bm{h}_{i}^{T}\sum_{k=1}^{K}\mathbf{v}_{k}e^{j(\phi_{k}-\phi_{i})}\right)=0}\ ,\ \forall {i}\\
    &&& \text{Re}{\left(\bm{h}_{i}^{T}\sum_{k=1}^{K}\mathbf{v}_{k}e^{j(\phi_{k}-\phi_{i})}\right)\geq \sqrt{\Gamma_{i}N_{0}}} \ , \ \forall {i}.
    \end{aligned}
\end{equation}

\section{Learning-Based SLP for Power minimization  problem}\label{non_robust_prob}
This section presents detailed formulations of a learning-based CI power minimization problem for strict phase angle rotation assuming a perfect channel state information (CSI) known at the BS. The power minimization problem for the case where the phase angle of the interfering symbols strictly aligns with the angle of the symbols of interest. If the maximum angle shift in the constructive interference region is zero, i.e., all the interfering signals completely overlap on the signal of interest ($\varphi=0$, see \cite{li2020tutorial} for details).

It is often difficult to derive the closed-form solution to problem $\mathcal{}{P}_{\text{st}}$ due to the in-equality constraints. Therefore, conventional iterative solvers are usually used to find sub-optimal solutions. Motivated by the recent adoption of an IPM for image restoration \cite{bertocchi2020deep}, we propose an unsupervised learning framework that unfolds a constrained optimization problem into a sequence of neural network layers for a multi-user MIMO beamforming. We first convert (\ref{p1}) to a general form of proximal IPM. The measure of the fidelity of the solution to (\ref{p1}) is determined by learning a set of penalty parameters in the form of Lagrange multipliers associated with the constraints. We define the channel vector based on (\ref{p1}) as follows
\begin{equation} \label{h_hat}
     \tilde{\mathbf{h}}_{i}=\mathbf{h}_{i}\sum_{k=1}^{K}e^{j(\phi_{k}-\phi_{i})}
\end{equation}  
 
\begin{equation} \label{w}
     {\mathbf{v}}=\sum_{k=1}^{K}\mathbf{v}_{k}.
\end{equation}
From (\ref{h_hat}) and (\ref{w}), we have
\begin{equation}\label{pro_comp}
    \tilde{\mathbf{h}}_{i}\mathbf{v}=(\tilde{\mathbf{h}}_{Ri}+j\tilde{\mathbf{h}}_{Ii})(\mathbf{v}_{R}+j\mathbf{v}_{I})
\end{equation}
where $\tilde{\mathbf{h}}_{R}=\text{Re}(\tilde{\mathbf{h}}_{i})$, $\tilde{\mathbf{h}}_{I}=\text{Im}(\tilde{\mathbf{h}}_{i})$, $\mathbf{v}_{R}=\text{Re}(\mathbf{v})$ and $\mathbf{v}_{I}=\text{Im}(\mathbf{v})$. Let $\boldsymbol{\Upsilon} =[ \tilde{\mathbf{h}}_{R}\ ; \ \tilde{\mathbf{h}}_{I}]$, $\mathbf{v}_{1}=[
\mathbf{v}_{R}\ -\mathbf{v}_{I}]^{T}$ and $\mathbf{v}_{2}=[\mathbf{v}_{I}\ \ \mathbf{v}_{R}]^{T}$. To simplify the analysis, we partition the complex vectors into the real and imaginary parts as follows:
$\text{Re}(\tilde{\mathbf{h}}_{i}^{T}\mathbf{v})=\boldsymbol{\Upsilon}_{i}^{T}\mathbf{v}_{1}$ and $\text{Im}(\tilde{\mathbf{h}}_{i}^{T}\mathbf{v})= \boldsymbol{\Upsilon}_{i}^{T}\boldsymbol{\Omega}\mathbf{v}_{1}$,
where 
\begin{equation}
{\mathbf{v}_{2}=\boldsymbol{\Omega}}\mathbf{v}_{1}\ \text{and}\ \boldsymbol{\Omega}=\begin{bmatrix}
\mathbf{O}_{N} & -\mathbf{I}_{N}\\
\mathbf{I}_{N} & \mathbf{O}_{N}
\end{bmatrix}; \ \in \mathbb{R}^{2N \times 2N}.
\end{equation}
Therefore, the multicast equivalent of (\ref{p1}) is 
\begin{equation} \label{p1_st}
    \begin{aligned}
    & \underset{\mathbf{\{v_{1}\}}}{\text{min}}
    & & {\norm{\mathbf{v}_{1}}^2} \\
    & \text{s.t.}
    & & {\boldsymbol{\Upsilon}_{i}^{T}\boldsymbol{\Omega}\mathbf{v}_{1}=0}\ ,\ \forall {i}\\
    &&& \boldsymbol{\Upsilon}_{i}^{T}\mathbf{v}_{1}\geq \sqrt{\Gamma_{i}N_{0}} \ , \ \forall {i}.
    \end{aligned}
    \end{equation}
\subsection{Interior Point Method (IPM)}
Consider a general form of a nonlinear constrained optimization of the form \cite{hauser2007interior}:   
\begin{equation}
    \begin{aligned}
    & \underset{\mathbf{x \in{\mathbb{R}^{N}}}}{\text{min}}
    & & {f(\mathbf{z})} \\
    & \text{s.t.}
    & & \mathcal{C(\mathbf{z})}=0\\
    &&& \mathbf{z}\geq 0.
    \end{aligned}
\end{equation}
 
The reason for adopting IPM is to replace the initial constrained optimization problem with a chain of unconstrained sub-problems of the form:
\begin{equation}\label{sub_prob0}
    \begin{aligned}
    & \underset{\mathbf{x \in{\mathbb{R}^{N}}}}{\text{min}} f(\mathbf{z})+\lambda{\mathbcal{D(\mathbf{z})}}+\mu{\mathbcal{B(\mathbf{z})}}.
    \end{aligned}
\end{equation}
where $\mathbcal{B}$ is the logarithmic barrier function associated with inequality constraint with unbounded derivative at the boundary of the feasible domain, $\mathbcal{D}$ is associated with equality constraint, $\mu$ and $\lambda$ are the Lagrangian multipliers for inequality and equality constraints, respectively.\par
To facilitate the solution of (\ref{p1_st}), we introduce additional notations. For every inequality constraint, $\gamma \in \{0,+\infty\}$ and $\mathbf{v}_{1} \in \mathbb{R}^{2N\times1}$, we define the proximity operator as in \cite{hauser2007interior} with respect to (\ref{sub_prob0}), which we shall later use to compute the projected gradient descent as 
\begin{equation}\label{prox_op}
   \begin{aligned}
   \text{prox}_{\gamma\mathcal{G}}{(\mathbf{v}_{1})}= & \underset{\mathbf{v_{1} \in{\mathbb{R}^{2N\times1}}}}{\text{argmin}}
    & & {\frac{1}{2}\norm{\mathbf{v}_{0}-\mathbf{v}_{1}}}_{2}^{2}+\gamma\mathcal{G}({\mathbf{v}_{1}})
    \end{aligned},
\end{equation}
where $\gamma$ is the step-size taken for computing the gradients of the objective function, $\mathcal{G}$ is the function that defines the barrier operator and $\mathbf{v}_{0}$ is the initial value of the precoding vector.\par

To convert (\ref{p1_st}) into its equivalent barrier function problem, we get raid of the inequality constrain and translate it into a barrier term of the form \cite{pustelnik2017proximity}
\begin{equation}\label{b1}
    \begin{aligned}
    & \underset{\mathbf{x_{i} \in{\mathbb{R}^{n}}}}{\text{min}}
    & & {f(\mathbf{z})}-\mu\sum_{i=1}^{p}\ln(\mathbf{z_{i})} \\
    & \text{s.t.}
    & & \mathcal{C(\mathbf{z})}=0.
    \end{aligned}
\end{equation}

\subsubsection{Affine Constraints}\label{Affine_const}
Consider a half-space constraint expressed as \cite{bertocchi2020deep}:
\begin{equation}\label{affine}
    \mathbcal{C}=\{\mathbf{z}\in\mathbb{R}^{N}\lvert{b}^{T}\mathbf{z}\leq{c}\}
\end{equation}

As shown in \cite{bertocchi2020deep}, the $\mathbcal{B}$ function associated to (\ref{affine}) is defined as
 \begin{equation}\label{prox_func2}
  \mathbcal{B(\mathbf{z})} = \left \{
  \begin{aligned}
    &-\ln{\left(c-b^{T}\mathbf{z}\right)}, && \text{if}\ b^{T}\mathbf{z}<c,\ \forall \ \mathbf{z}\in\mathbb{R}^{N} \\
    &+\infty, && \text{otherwise}
  \end{aligned} \right.
\end{equation}

Following (\ref{b1}), we can express (\ref{p1_st}) as
\begin{equation}\label{b2}
    \begin{aligned}
    & \underset{\mathbf{w_{1}}}{\text{min}}
    & & {f(\mathbf{v}_{1})}-\mu\sum_{i=1}^{p}{\ln\left(\boldsymbol{\Upsilon}_{i}^{T}\mathbf{v}_{1}-\sqrt{\Gamma_{i}N_{0}}\right)} \\
    & \text{s.t.}
    & & {\boldsymbol{\Upsilon_{i}^{T}\Omega}{\mathbf{v}_{1}}=0}.
    \end{aligned}
\end{equation}
For all $\mu >0$, $\lambda >0$ and $\mathbf{v}_{1}$, we define $\mathbcal{B}$ as in (\ref{prox_func2}), so that the proximity operator can be defined as follows
\begin{equation}
    \Phi(\mathbf{v}_{1},\gamma,\mu)=\text{prox}_{\gamma\mu\boldsymbol{\mathbcal{B}}}{(\mathbf{v}_{1})}.
\end{equation}

In what follows, we provide the expression of $\Phi$ and its corresponding derivatives with respect to the optimization variable $\mathbf{v}_{1}$, the step-size and the barrier parameters $(\gamma, \mu)$ for affine constraint, which will be used for training the neural network using a gradient backpropagation algorithm. Finally, following the above formulations, the proximal barrier function for the strict phase rotation is reduced to the following expression
\begin{equation}\label{prox_func4}
  \mathbcal{B(\mathbf{v}_{1})} = \left \{
  \begin{aligned}
    &-\ln{\left(\boldsymbol{\Upsilon}_{i}^{T}\mathbf{v}_{1}-\sqrt{{\Gamma}_{i} N_{0}}\right)}, && \text{if}\ \boldsymbol{\Upsilon}_{i}^{T}\mathbf{v}_{1} \geq\sqrt{{\Gamma}_{i}N_{0}} \\
    &+\infty, && \text{otherwise.}
  \end{aligned} \right.
\end{equation} 
It can be easily shown that for every precoding vector $\mathbf{v}_{1} \in\mathbb{R}^{2N\times1}$, the proximity operator of $\mu\gamma\mathbcal{B}$ at $\mathbf{v}_{1}$ is given by
\begin{multline}\label{prox_op_strct}
    \Phi(\mathbf{v}_{1},\mu,\gamma)=\mathbf{v}_{1}+ \\ \frac{\boldsymbol{\Upsilon_{i}}^{T}\mathbf{v}_{1}-\sqrt{\Gamma_{i}N_{0}}-\sqrt{(\boldsymbol{\Upsilon}_{i}^{T}\mathbf{v}_{1}-\sqrt{\Gamma_{i}N_{0}})^{2}+4\gamma\mu\norm{\boldsymbol{\Upsilon}_{i}^{T}}_{2}^{2}}}{{{2\norm{\boldsymbol{\Upsilon}_{i}}_{2}^{2}}}}\boldsymbol{\Upsilon}_{i}.
\end{multline}

Furthermore, the Jacobian matrix of $\Phi$ with respect to $\mathbf{v}_{1}$, and the derivatives of $\Phi$ with respect to $\gamma$ and $\mu$ are as follows
\begin{multline}\label{jacob_mat_strct}
\mathcal{J}_{\Phi}\mid_{(\mathbf{v}_{1})}=\mathbf{I}_{2N}+\frac{1}{2\norm{\boldsymbol{\Upsilon}_{i}}_{2}^{2}} \ \times \\ \left(1-\frac{\boldsymbol{\Upsilon}_{i}^{T}-\sqrt{\Gamma_{i}N_{0}}}{\sqrt{(\boldsymbol{\Upsilon}_{i}^{T}\mathbf{v}_{1}-\sqrt{\Gamma_{1}N_{0}})^{2}+4\gamma\mu\norm{\boldsymbol{\Upsilon}_{i}^{T}}_{2}^{2}}}\right)\boldsymbol{\Upsilon_{i}\Upsilon_{i}^{T}}
\end{multline}

\begin{equation}\label{ieq_const_strict}
    \Delta_{\Phi}\mid_{({\mu})}=\frac{-\gamma}{{\sqrt{(\boldsymbol{\Upsilon}_{i}^{T}\mathbf{v}_{1}-\sqrt{\Gamma_{1}N_{0}})^{2}+4\gamma\mu\norm{\boldsymbol{\Upsilon}_{i}^{T}}_{2}^{2}}}}\boldsymbol{\Upsilon_{i}}
\end{equation}

\begin{equation}\label{eq_const_strict}
    \Delta_{\Phi}\mid_{({\gamma})}=\frac{-\mu}{{\sqrt{(\boldsymbol{\Upsilon}_{i}^{T}\mathbf{v}_{1}-\sqrt{\Gamma_{1}N_{0}})^{2}+4\gamma\mu\norm{\boldsymbol{\Upsilon}_{i}^{T}}_{2}^{2}}}}\boldsymbol{\Upsilon_{i}}
\end{equation}
where $\mathbf{I} \in \mathbb{R}^{2(N \times N)}$ is identity matrix.\par
Finally, the learning algorithm for every update rule is thus the unfolded (\ref{p1_st}) as a sequence of sub-problems with respect to the constraints as follows
\begin{equation}\label{prox_func}
    \begin{aligned}
    & \underset{\mathbf{v_{1} \in{\mathbb{R}}^{2N\times1}}}{\text{min}}
    & & {\norm{\mathbf{v}_{1}}}_{2}^{2}+\lambda(\boldsymbol{\Upsilon_{i}^{T}\Omega}{\mathbf{v}_{1}}) +\mu{\mathbcal{B(\mathbf{v}_{1})}}.
    \end{aligned}
\end{equation}\par
Using the proximity operator of the barrier, the update rule for every iteration is given by
\begin{equation}\label{beam_update}
\mathbf{v}_{1}^{[r+1]}=\text{prox}_{\gamma^{[r]}\mu^{[r]}\mathcal{B}}\left(\mathbf{v}_{1}^{[r]}-\gamma^{[r]}\Delta{\mathbcal{E}(\mathbf{v}_{1}^{[r]},\lambda^{[r]})}\right),
\end{equation}
where 
\begin{equation}\label{eq_constr_func}
    \mathbcal{E}(\mathbf{v}_{1}^{[r]},\lambda^{[r]}) = {\norm{\mathbf{v}_{1}}}_{2}^{2}+\lambda(\boldsymbol{\Upsilon_{i}^{T}\Omega}{\mathbf{v}_{1}}).
\end{equation}

The update function can thus be expressed as 
\begin{multline}
\mathbcal{H}(\mathbf{v}_{1}^{[r]},\gamma^{[r]},\mu^{[r]},\lambda^{[r]})=\\ \text{prox}_{\gamma^{[r]}\mu^{[r]}\mathcal{B}}\left(\mathbf{v}_{1}^{[r]}-\gamma^{[r]}\Delta{\mathbcal{E}(\mathbf{v}_{1}^{[r]},\lambda^{[r]})}\right),
\end{multline}
and $\Delta=\frac{\partial{\mathbcal{E}(\mathbf{v}_{1}^{[r]},\lambda^{[r]})}}{\partial{\mathbf{v}_{1}^{[r]}}}$.

\subsubsection{Duality and Loss Function for the Strict Phase Formulation}
Since we are interested in learning the optimal solution via unsupervised learning (without target labels), we firstly formulate a primal-dual problem. This formulation is then used to derive the optimization variable (precoding vector) as a function of dual variables (Lagrangian multipliers) associated with the constraints. The Lagrangian function can be expressed as
\begin{multline}\label{Lag_strict}
\mathcal{L}_\text{st}(\mathbf{v}_{1} ,\lambda ,\ \mu ) =\Vert \mathbf{v}_{1}\Vert_{2}^{2} +\sum ^{K}_{k=1} \lambda _{k}\boldsymbol{\Upsilon }^{T}_{i}\boldsymbol{\Omega }\mathbf{v}_{1} + \\ \sum ^{K}_{k=1} \mu _{k}\left(\sqrt{\Gamma _{i} N_{0}} -\boldsymbol{\Upsilon }^{T}_{i}\mathbf{v}_{1}\right).
\end{multline}

The optimal precoder can be found by minimizing (\ref{Lag_strict}) with respect to $\mathbf{v}_{1}$ (differentiating $\mathcal{L}_\text{st}(\cdotp)$ w.r.t $\mathbf{v}_{1}$). The optimal precoder is thus
\begin{equation}\label{optima_prec_str}
\mathbf{v}_{1}=\frac{\boldsymbol{\mu}^{T}\cdotp\boldsymbol{\Upsilon}_{i}-\boldsymbol{\lambda}^{T} \cdotp \boldsymbol{\Omega}\boldsymbol{\Upsilon}_{i}}{2}.
\end{equation}
The above expression in (\ref{optima_prec_str}) is used to generate the training input (precoding vector) by initializing the Lagrangian multipliers ($\lambda$ and $\mu$) randomly and then train the neural network to learn their best values that minimize the loss function (Lagrangian function). The loss function is modified by adding $\mathbcal{l}2$-norm regularization over the weights to adjust the learning coefficients to stabilize the learning process. The loss function over $B$ training batches is finally expressed as 
\begin{multline}\label{Loss_strict}
\mathcal{L}_\text{st}(\mathbf{v}_{1} ,\lambda ,\ \mu ) =\frac{1}{B}\sum^{B}_{i=1}\left(\Vert \mathbf{v}_{1}\Vert_{2} ^{2} + \boldsymbol{\lambda}\boldsymbol{\Upsilon }^{T}_{i}\boldsymbol{\Omega }\mathbf{v}_{1}\right) + \\ \frac{1}{B}\sum^{B}_{i=1}\left(\boldsymbol{\mu}\left(\sqrt{\Gamma _{i} N_{0}} -\boldsymbol{\Upsilon }^{T}_{i}\mathbf{v}_{1}\right)\right)+ 
\frac{\vartheta}{BL}\sum^{B}_{i=1}\sum_{i=1}^{L}\Vert \boldsymbol{\theta}_{i}\Vert_{2} ^{2},
\end{multline}
where $\boldsymbol{\theta}$ is the learning parameter associated with the weights and $\vartheta >0$ is the penalty parameter that controls the bias and variance of the learning coefficients, $B$ and $L$ are training batch size (number of channel realization) and the number of layers respectively. 

\begin{figure*}[!t]
    \centering
    \includegraphics[width=7.2in,height=3.2in]{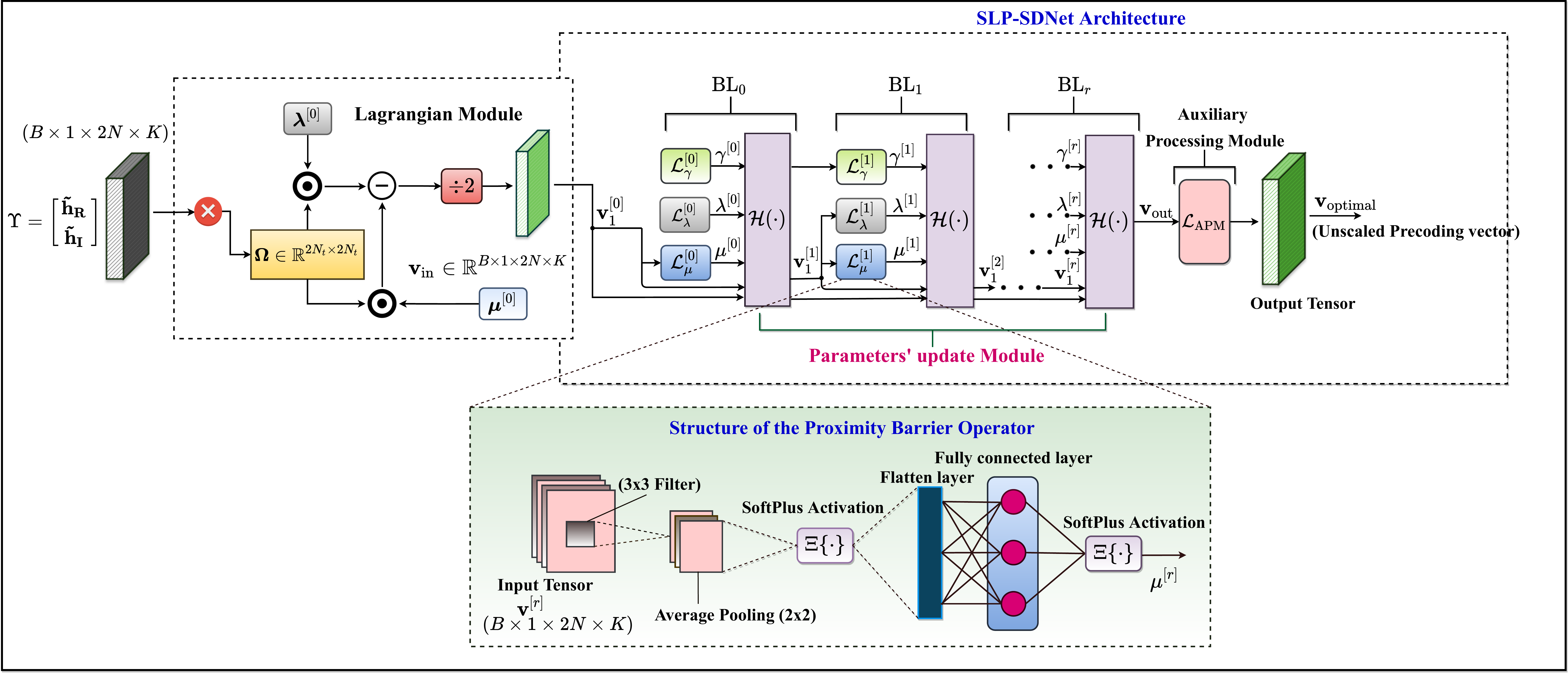}
    \caption{Complete SLP-SDNet Architecture showing the internal structure of the Barrier Operator.}
    \label{fig:DNNBF_Arch}
\end{figure*}

\subsection{Deep Proximal Strict Symbol-Level Precoding Network (SLP-SDNet)}
The optimization problem is unfolded over \textit{r-th} iterations, and the Lagrange multiplier associated with the equality constraint is wired across the network to provide additional flexibility \cite{bertocchi2020deep}. The unfolded neural network is trained in an unsupervised fashion without target labels. We build the structure of the learning framework based on (\ref{beam_update}) and the algorithm presented in \cite{bertocchi2020deep}, which gives rise to Algorithm \ref{algorithm_1}.

\begin{algorithm}
 \caption{Feed-forward-Backward Proximal IPM}
 \begin{algorithmic}[1]\label{algorithm_1}
 \renewcommand{\algorithmicrequire}{\textbf{Input:}}
 \renewcommand{\algorithmicensure}{\textbf{Output:}}
 \REQUIRE $\mathbf{v}_{1}^{[0]}$, ${\gamma}^{[0]}$, ${\boldsymbol{\lambda}}^{[0]}$ and $\boldsymbol{\mu}^{[0]}$
 \ENSURE  $\mathbf{v}_{1}$ 
 \\ \textit{Initialization} :
  \STATE randomly initialize $\mathbf{v}_{1}^{[0]}\in{\mathbb{R}^{2N\times1}}$, $\boldsymbol{\mu}^{[0]}>0$, $\boldsymbol{\lambda}^{[0]}>0$ and $\gamma^{[0]}>0$ $\forall\ {i}=1,\ \cdots,\ K$ 
 \\ \textit{Loop over r-th iterations}
  \FOR {$r=0$ to $L$}
  \STATE $\mathbf{v}_{1}^{[r+1]}=\text{prox}_{\gamma^{[r]}\mu^{[r]}\mathcal{B}}\left(\mathbf{v}_{1}^{[r]}-\gamma^{[r]}\Delta{\mathbcal{E}(\mathbf{v}_{1}^{[r]},\lambda^{[r]})}\right).$
  \ENDFOR
 \RETURN $\mathbf{v}_{1}$ 
 \end{algorithmic} 
\end{algorithm}

\begin{algorithm}
 \caption{Proximity Barrier Operator for Strict phase rotation}
 \begin{algorithmic}[1]\label{algorithm_strict}
 \renewcommand{\algorithmicrequire}{\textbf{Input:}}
 \renewcommand{\algorithmicensure}{\textbf{Output:}}
 \REQUIRE $\mathbf{h}_\text{{R}}$, $\mathbf{h}_\text{{I}}$, $\Gamma_{i}$ and ${N}_{0}\ (\text{noise power})$
 \ENSURE  $\mathbf{v}_{1}$, $\gamma$, $\boldsymbol{\mu}$ and $\boldsymbol{\lambda}$
 \\ \textit{Initialization} :
  \STATE randomly initialize $\mathbf{v}_{0}\in{\mathbb{R}^{2N\times1}}$, $\boldsymbol{\mu}^{[0]}>0$, $\boldsymbol{\lambda}^{[0]}>0$ and $\gamma^{[0]}>0$ $\forall\ {i}=1,\ \cdots,\ K$.
  \STATE \text{Compute the Barrier function $\mathbcal{B(\mathbf{v}_{1})}$ using} \text{function (\ref{prox_func4})}.
  \STATE \text{Compute the Proximity Operator of the Barrier at} $\mathbf{v}_{0}$ \text{using (\ref{prox_op}), where  $\mathcal{G}=\mu\mathbcal{B(\mathbf{v}_{1})}$ }.
  \STATE \text{Compute the derivatives of the Proximity Operator} \text{w.r.t} $\mathbf{v}_{1}$, $\mu$ and $\gamma$ \text{using} (\ref{jacob_mat_strct}), (\ref{ieq_const_strict}) and (\ref{eq_const_strict}).
  \STATE {Update the training variables as follows:}
    \begin{enumerate}[label=(\alph*)]
      \item $\mu^{[r+1]}=\mu^{[r]}-\eta\frac{\partial{\Phi}}{\partial{\mu}}$
      \item  $\gamma^{[r+1]}=\gamma^{[r]}-\eta\frac{\partial{\Phi}}{\partial{\gamma}}$
      \item  $\lambda^{[r+1]}=\lambda^{[r]}-\eta      \frac{\partial{\mathbcal{E}(\mathbf{v}_{1}^{[r]},\lambda^{[r]})}}{\partial{\lambda^{[r]}}}$ using (\ref{eq_constr_func})
  \end{enumerate}
  \text{where} $\eta$ \text{is the learning rate}.
  \STATE \text{Use the results in step 5 and the Algorithm \ref{algorithm_1}} \text{to obtain the optimal precoding tensor}.
 \end{algorithmic} 
\end{algorithm}

For every \textit{r-th} iterations (r-th layer) $\mathcal{L}^{[r]}$, there exist three latent structures associated with the learnable parameters ($\mu$, $\gamma$ and $\mu$) $\mathcal{L}_{\mu}^{[r]}$, $\mathcal{L}_{\gamma}^{[r]}$ and $\mathcal{L}_{\lambda}^{[r]}$. As shown in Fig. \ref{fig:DNNBF_Arch}, each of these structures forms a learning block for computing the barrier parameter ($\mu$) associated with the inequality constraint, the step-size for update rule ($\gamma$) and finally ($\lambda$), which is related to the equality constraint and all of them must be positive. To impose such constraint, a \textit{`Softplus sign'} function is used. Hence, the step-size and the parameters associated with the constraints can all be estimated as ${\gamma}^{[r]}=\mathcal{L}_{\gamma}^{[r}= \text{Softplus}(\mathbf{z}^{[r]})$. The output of the last three hidden structures is connected to an auxiliary processing module (APM) to convert it into the required transmit precoding vector. The APM consists of 4 convolution layers and 3 activation layers, a \textit{``Batch Normalization''} layer placed between them. Therefore, the Proximal Barrier function for a strict phase formulation is summarized in Algorithm \ref{algorithm_strict}.

Finally, the output from the auxiliary processing block is the precoding vector in the real domain. The relation:\\ $\mathbf{v}_{1} = [
\mathbf{v}_{R} \ -\mathbf{v}_{I}]^{T}$  is used to convert it to its equivalent complex domain for every SINR value of the i-th user.

\subsection{The proposed Learning Structure and the general NN Architecture}
Using (\ref{beam_update}) and Algorithm \ref{algorithm_1}, we show a startling correlation between our scheme and the universal feed-forward DNN. Generally, an open-chained neural network (NN) structure can be derived from (\ref{beam_update}) as follows
\begin{equation}\label{beam_update2}
\mathbf{v}_{1}^{[r+1]}=\text{prox}_{\gamma^{[r]}\mu^{[r]}\mathcal{B}}\left[\left(\mathbf{I}_{2N}-2\gamma^{[r]}\right)\mathbf{v}_{1}^{[r]}+\lambda^{[r]}\Upsilon_{i}\Omega\right].
\end{equation}\par
By letting  
$\mathbf{W}_{r}=\mathbf{I}_{2N}-2\gamma^{[r]}$, $\mathbf{b}_{r}=\lambda^{[r]}\Upsilon_{i}\Omega$ and $\boldsymbol{\Pi}_{r}=\text{prox}_{\gamma^{[r]}\mu^{[r]}\mathcal{B}}$, the r-layer network $\mathcal{L}^{[r-1]}\cdots\mathcal{L}^{[0]}$ will correspond to the following
\begin{multline}\label{neural_net}
  \boldsymbol{\Pi}_{R-1}\left(\mathbf{W}_{R-1}\mathbf{v}_{1}^{[R-1]}+\mathbf{b}_{R-]}\right),\cdots,\ \boldsymbol{\Pi}_{0}\left(\mathbf{W}_{0}\mathbf{v}_{1}^{[0]}+\mathbf{b}_{0}\right) \\
  \forall\ 0\leq{r}\leq{R-1} \in \text{R-layers}, 
\end{multline}
where $[\mathbf{W}_{r}]_{0\leq{r}\leq{R-1} }$ and $[\mathbf{b}_{r}]_{0\leq{r}\leq{R-1}}$ are described as weight and bias parameters respectively. The identity square matrix is defined as $ \mathbf{I}_{2N} \in \mathbb{R}^{2(N\times N)}$. The nonlinear activation functions are defined by $[\boldsymbol{\Pi}_{r}]_{0\leq{r}\leq{R-1}}$ and can be obtained from the proximal operator. Furthermore, $\boldsymbol{\Pi}_{r}$ can be expressed as sum of a bias and a proximal activation operator.
\subsection{SLP-SDNet Training and Testing}\label{training}
The SLP-SDNet has two modules; the parameter module and the auxiliary module. The parameter module consists of three structures associated with Lagrangian multipliers (equality and inequality constraints) and the training step-size. The proximity barrier function is related to the inequality constraint and forms the parameter module. It is constructed with one convolutional layer, an average pooling layer, a fully connected layer, and a softPlus layer so that the output is constrained to a positive real value. The parameter update module contains \textit{r}-th blocks and is trained block-wise for \textit{l}-th number of iterations. Similarly, the auxiliary unit is trained for \textit{k}-th iterations. It is important to note that the number of training iterations of the parameter update module may not necessarily be equal to that of the auxiliary unit. We train the parameter update unit with 15 iterations and the auxiliary unit for 10 iterations. During the inference, a feed-forward pass is performed over the whole layers using the learned Lagrangian multipliers to calculate the precoding vector using (\ref{optima_prec_str}). The trained model is run over different SINR values to output the optimal precoding matrix. 
\section{Results and Discussion}\label{results}
\subsection{Simulation Setup}
We consider a downlink scenario, where the BS has four antennas ($N=4$) that serve $K$ single users, assuming a perfect known CSI. We generate 50,000 training samples and 2000 test samples of the channel coefficients randomly drawn from a normal distribution with zero mean and unit variance using (\ref{h_hat}). The datasets are normalized by the data symbol so that data entries are within the nominal range. The transmit data symbols are modulated using a QPSK and 8PSK modulation schemes; and the SINR is randomly generated from uniform distribution $\Gamma_{\text{train}} \sim \mathcal{U}(\Gamma_\text{low},\ \Gamma_\text{high})$. A stochastic gradient descent algorithm with Adam optimizer is used to minimize the Lagrangian function (loss function). For every training iteration, the learning rate is reduced by $\beta=0.65$ to help the learning algorithm converge faster. The implementation is done on Pytorch 1.7.1 and Python 3.7.8 on a computer with the following specifications: Intel(R) Core (TM) i7-6700 CPU Core, 32.0GB RAM.

\begin{figure}[!t]
    \centering
        \centering
        \includegraphics[width=2.6in,height=2.2in]{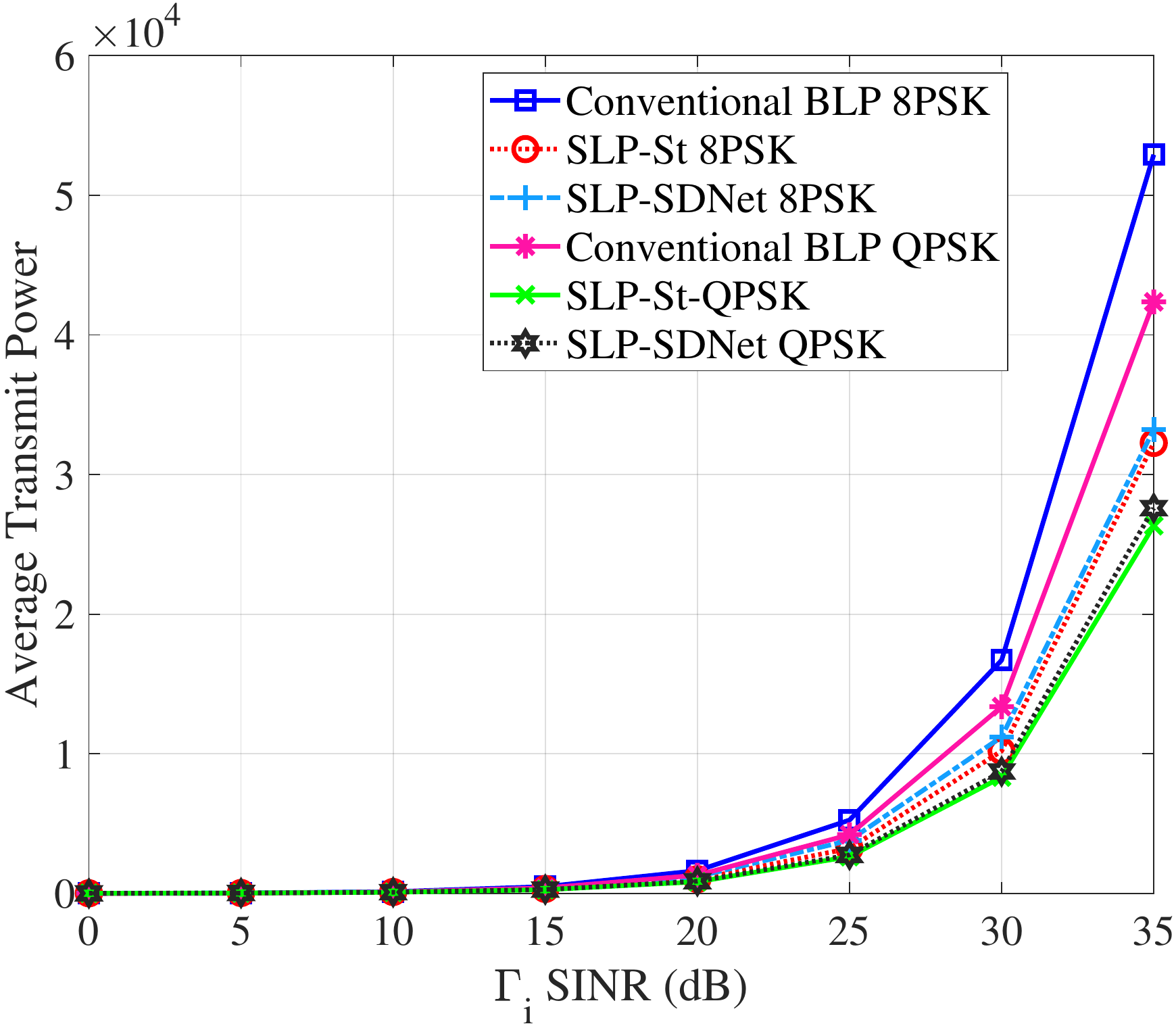}
        \caption{Transmit Power vs SINR averaged over 2000 test samples for Conventional BLP, SLP-St optimization-based and SLP-SDNet schemes, $N=4$, $K=4$.}
        \label{fig:FP_Nonrobust}
\end{figure}

\begin{figure}[!t]
    \centering
        \centering
        \includegraphics[width=2.6in,height=2.2in]{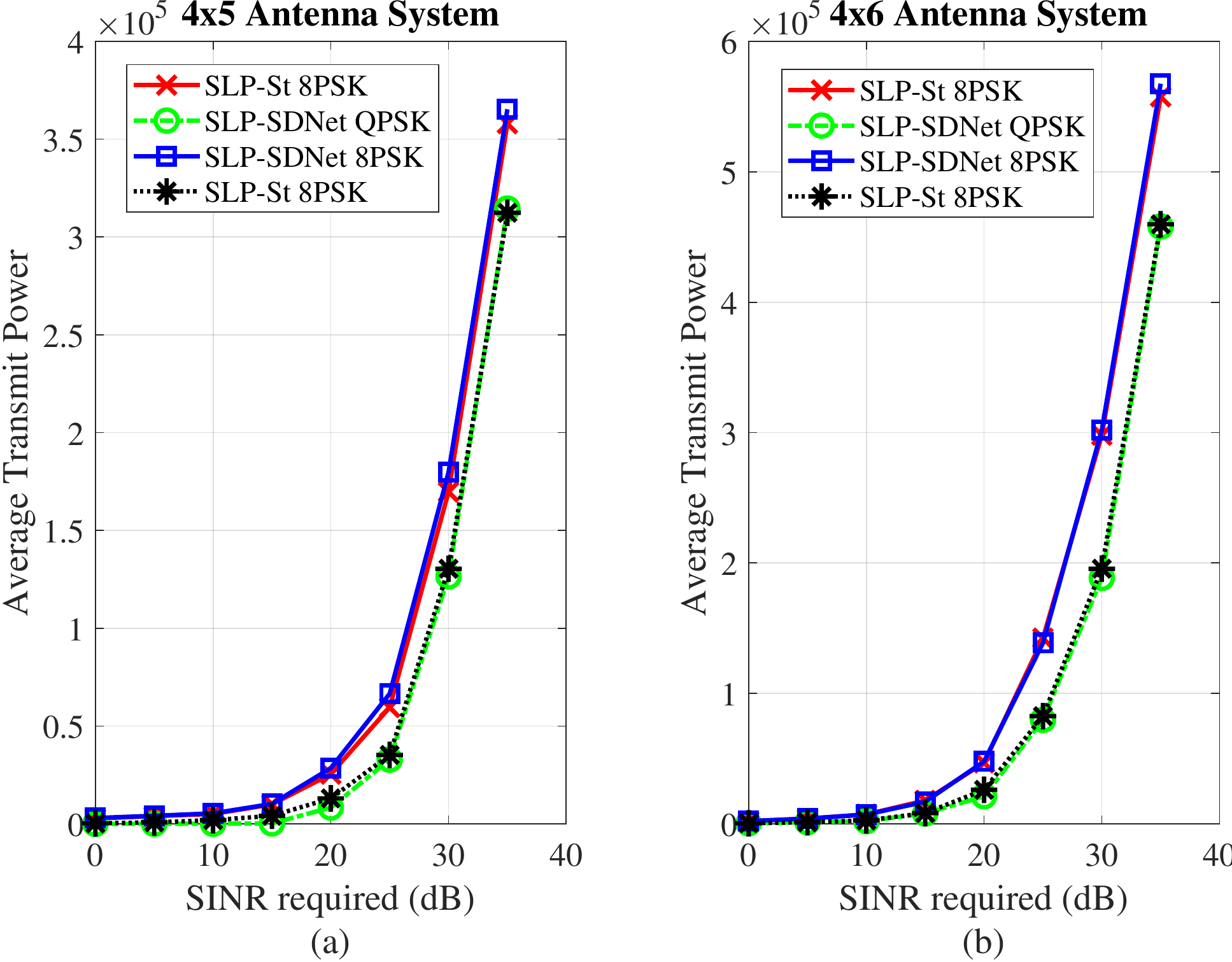}
        \caption{Transmit Power vs SINR averaged over 2000 test samples for SLP-St optimization-based and SLP-SDNet schemes for varying number of user; $N=4$, $K=5$ and $K=6$.}
        \label{fig:FP_Nonrobust2}
\end{figure}

\subsection{Performance Evaluation of SLP-SDNet}\label{nonrobust_performance}
We consider a SLP-SDNet for strict phase angle rotation problems (\ref{Loss_strict}). Our proposed unsupervised learning framework's performance is evaluated against the benchmark precoding designs in \cite{bjornson2014optimal, masouros2015exploiting}. We compare the average transmit power of the conventional BLP approach (\ref{conv_power_min}), the SLP-based problems (\ref{p1}), the proposed SLP learning-based precoding scheme based on Algorithm \ref{algorithm_strict}. Fig. \ref{fig:FP_Nonrobust} shows that the SLP-SDNet gives less transmit power than the conventional BLP scheme because, for a $4\times4$ system, there is inadequately available transmit power at the BS. It is also essential to note that the transmit power given by an SLP-SDNet is the same as for an SLP optimization-based solution at $SINR$ values below $30${dB}. However, the transmit power increases by 8\% for an SLP-SDNet solution over SLP optimization-based approach at $SINR$ greater than $30${dB}.\par 
Figs. \ref{fig:FP_Nonrobust2}(a) and \ref{fig:FP_Nonrobust2}(b) show the average transmit power for a given BS antennas ($N=4$) and varying number of users (5 and 6 users). We find from our simulation that while conventional BLP is only feasible for $K\leq M$, both SLP optimization-based algorithm and the proposed learning schemes are viable for all sets of $N$ BS antennas and $K$ mobile users. Furthermore, we also observe that the performance gap between the SLP optimization-based and proposed learning-based schemes closes as more users are served.

\subsection{Complexity Evaluation}
\begin{figure}[!t]
    \centering
    \includegraphics[width=2.6in,height=2.2in]{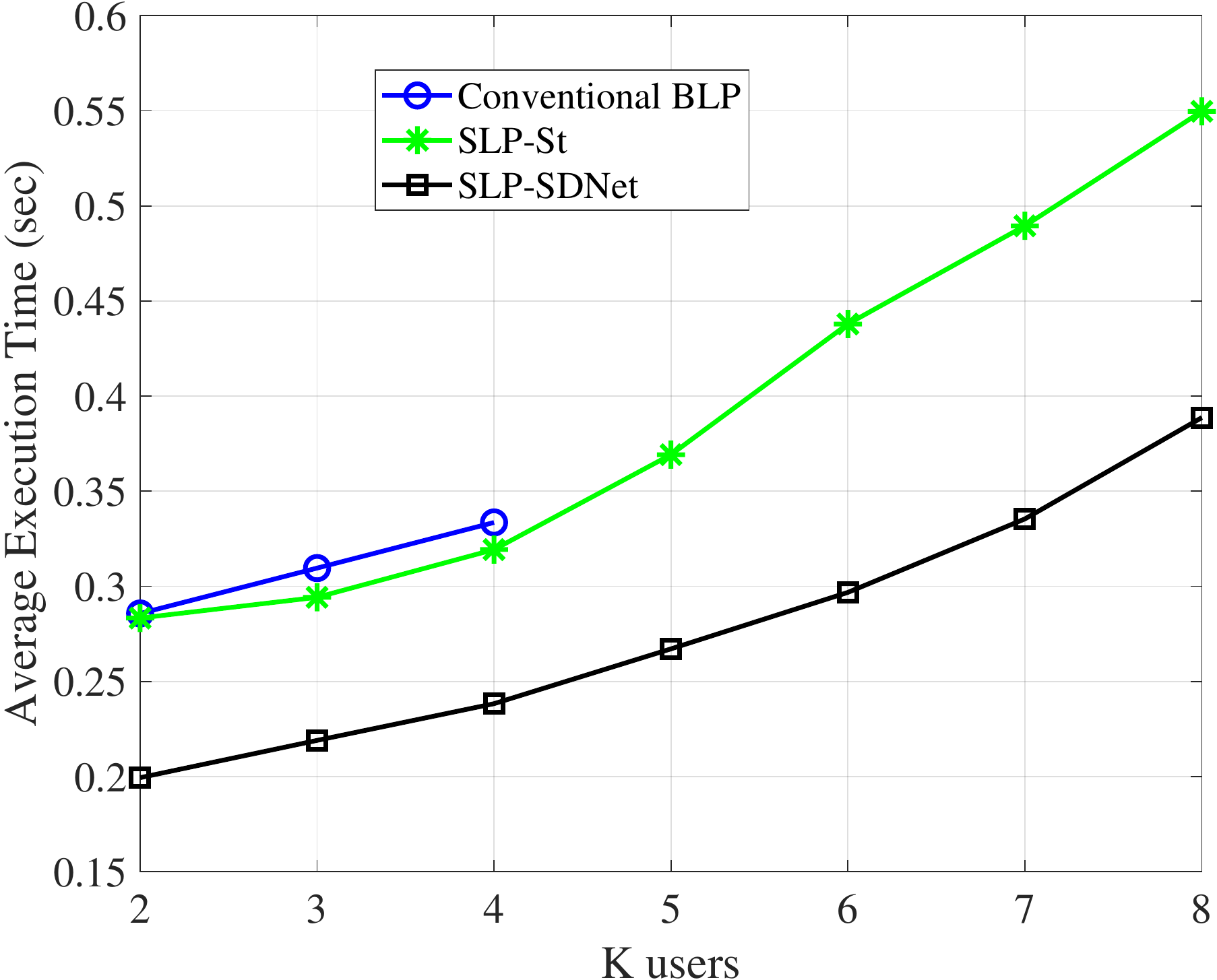} 
    \caption{Comparison of average execution time per sample averaged over 2000 test samples for Conventional BLP, SLP-St optimization-based and SLP-SDNet schemes.}
        \label{fig:EXECUTION_TIME}
\end{figure}

For a fair comparison, we measure the complexities of our proposals and the benchmark optimization-based precoding schemes in terms of the optimization algorithms' average execution time, as shown in Figs. \ref{fig:EXECUTION_TIME}. We observe that the average execution time of the SLP-SDNet scheme per symbol averaged over 2000 test samples offers $2\times$ decrease in execution time per data symbol because the predominant operations in SLP-SDNet during online training are matrix-matrix or vector-matrix convolution.  This shows that the proposed unsupervised learning-based precoding scheme offers a desirable trade-off between performance and computational complexity.
\section{Conclusion}\label{conclusion}
This paper proposes a fast unsupervised learning-based precoding framework for a multi-user downlink MISO system. The proposed learning technique exploits the constructive interference for the power minimization problem so that for given QoS constraints, the transmit power available for transmission is minimized. We use domain knowledge to develop an unsupervised learning architecture by unfolding the proximal interior point method barrier \textit{`log} function. Proximal barrier function for strict phase rotation is derived based on the nature and characteristics of the inequality constraints. 

\bibliographystyle{IEEEtran}
\bibliography{ref}

\end{document}